\documentstyle[sprocl,epsfig]{article}

\def\be{\begin{equation}}
\def\ee{\end{equation}}
\def\bea{\begin{eqnarray}}
\def\eea{\end{eqnarray}}

\begin{document}

\title{BL Lacertae: the multiwavelength  campaign of 2000}

\author{M. Ravasio, G. Tagliaferri, Ghisellini G.}

\address{Osservatorio Astronomico di Brera, Via Bianchi 46,\\
 I--23807 Merate, Italy\\E-mail: ravasio@merate.mi.astro.it} 

\author{M. B\"{o}ttcher}

\address{Department of Physics and Astronomy, Rice University,\\
 MS 108, 6100 S. Man street\\
Houston, TX 77005-1892, USA}

\author{H.D. Aller, M.F. Aller}

\address{Astronomy Department, University of Michigan, \\
Ann Arbor, MI 48109, USA}

\author{O. Mang}

\address{Institut f\"{u}r Experimentelle und Angewandte Physik, \\
Universit\"{a}t Kiel, Leibnizstra$\beta$e 15--19,\\
24118 Kiel, Germany}

\author{L. Maraschi}

\address{Osservatorio Astronomico di Brera, Via Brera 28, \\
I--20121 Milano, Italy}

\author{E. Massaro}

\address{Dipartimento di Fisica, Universit\'a La Sapienza, \\
P.le A. Moro 2, I--00185 Roma, Italy}

\author{C. M. Raiteri, M. Villata}

\address{Osservatorio Astronomico di Torino, Via Osservatorio 20,\\
10025 Pino Torinese, Italy}

\author{H. Ter\"{a}sranta}

\address{Mets\"{a}hovi Radio Observatory, Helsinki University of Technology,\\
Mets\"{a}hovintie 114, 02540 Kylm\"{a}l\"{a}, Finland}


\maketitle\abstracts{We present two {\it Beppo}SAX observations of BL Lacertae
as part of a multiwavelength radio--to--TeV campaign. 
During the first observation
we observe a faint Compton spectrum, while during the second, we detect 
a synchrotron spectrum with the highest [2-10] keV flux ever measured; 
above 10 keV an inverse Compton component begin to dominate. The synchrotron 
flux is very variable with time scales of $\sim 1$ hr.
We describe four different SED shifting the synchrotron peak both 
in frequency and flux  intensity 
and we sketch a scenario in which a blob moves along a jet
and can be  located in or outside the BLR. 
This implies different radiative mechanism at work, SSC or external Compton,
producing different high energy spectra.}
\section{Introduction}
Blazars are radio loud active galactic nuclei characterized by 
an extremely wide spectral range, from radio to $\gamma$--ray 
(sometimes up to TeV frequencies) and by fast and large variability:
simultaneous multiwavelength observations are therefore the 
most powerful tool to reveal the underlying mechanisms.
During the last 20 years, BL Lacertae, the BL Lac prototype, 
  has been the target of many multiwavelength campaigns
 (Bregman et al., 1990; Kawai et al., 1991;
Sambruna et al., 1999; Madejski et al., 1999; Ravasio et al., 2002).\\
X--ray observations are particularly interesting for this source, 
because they have revealed the transition from 
 synchrotron, which is rapidly varying,  to the more quiet 
inverse Compton emission  (Ravasio et al., 2002).\\
Therefore during the second half of 2000 a new multiwavelength campaign 
was organized, ranging from radio to TeV energies: the X--ray band
was covered by  {\it Beppo}SAX, 
with a $10^{5}$ sec run in the core of the campaign (July 17 -- August 11;
Ravasio et al., in prep.)
and by RXTE, which assured 3 short pointings per week 
(Marscher et al., in prep.). 
Besides X--ray, the campaign comprised radio 
(4.8, 8, 14.5 GHz: Michigan Radio Astronomy Observatory;
22, 37 GHz: Mets\"{a}hovi Radio Telescope) , optical (WEBT collaboration, 
Villata et al., 2000) and VHE $\gamma$--ray observations (CAT; HEGRA).
Because of an increase in the activity  of the source (Villata et al., 2002),
 the campaign was prolonged until the end of 2000,
 with a second {\it Beppo}SAX observation started at the end of October.  
During the autumn, the HEGRA team 
accumulated  10.5 hrs on--source time and was
able  to set an upper limit of $25\%$ of the Crab flux above 0.7 TeV
(Mang et al., 2001). For a more detailed description of the campaign, 
we refer to B\"{o}ttcher et al. (in prep).\\
\section{X--ray observations}
\begin{figure*}[t]
\begin{center}
\vskip -0.7 true cm
\hbox to \textwidth{
\centerline{
\hskip 1 true cm
\vbox{\epsfig{figure=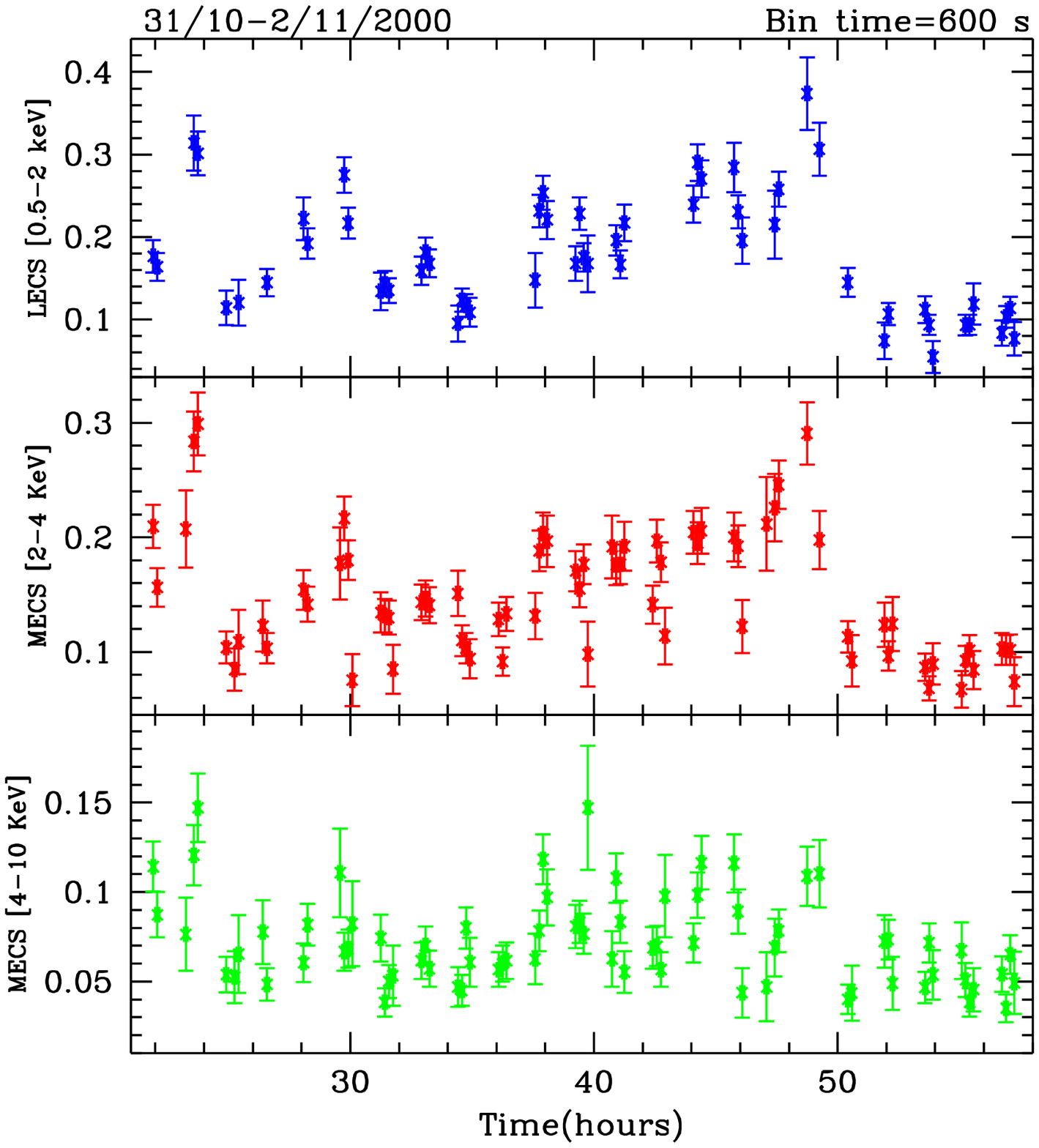, width=9cm}
}
\hskip -3 true cm
\vbox{\psfig{figure=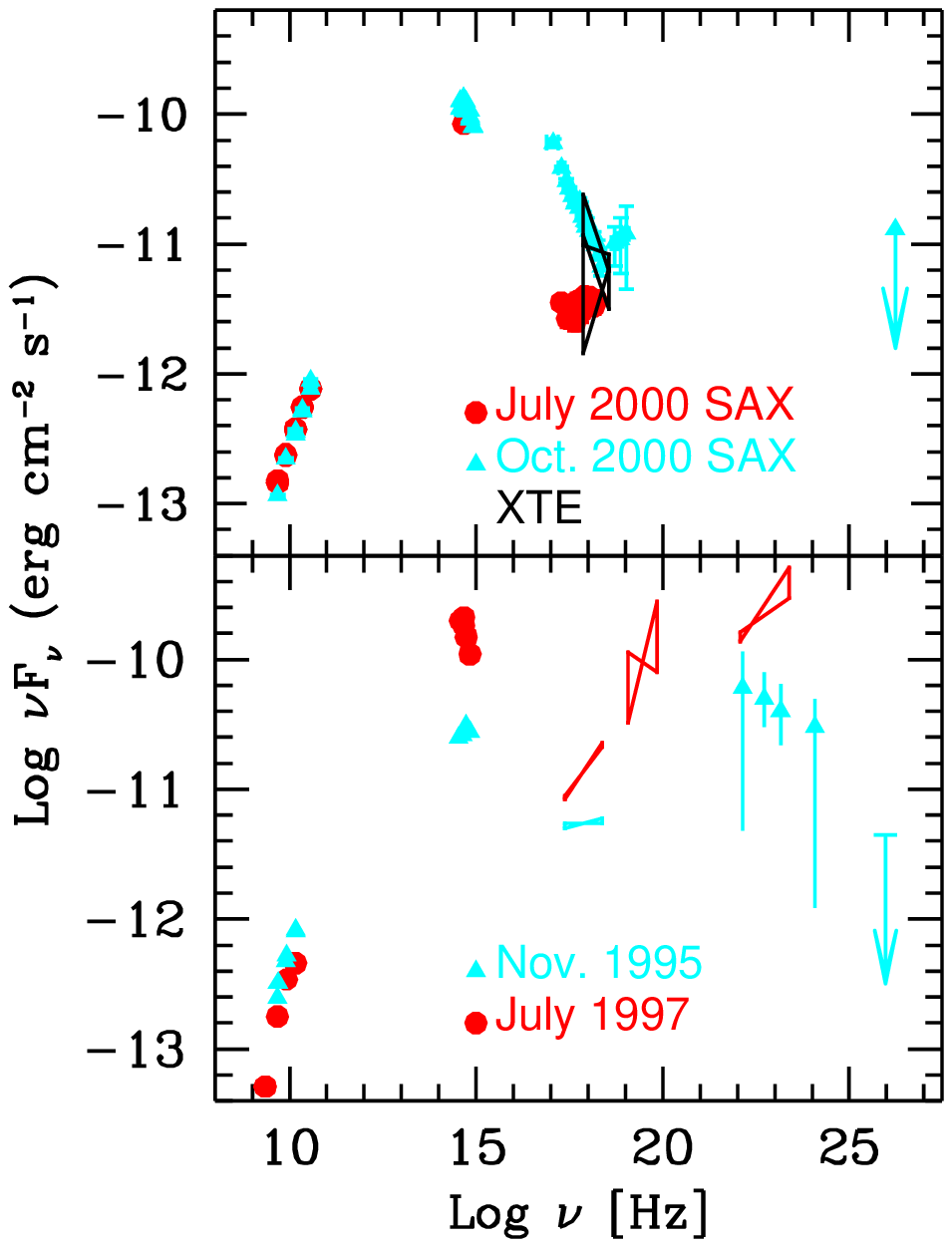, width=8cm}}
}
}
\vskip -0.3 true cm
\caption{Left: October--November 2000 {\it Beppo}SAX light curves.
Right top panel: simultaneous July 2000 and October--November 2000
observations. The arrow on the right represent the HEGRA upper limit, 
calculated at 0.7 TeV. The butterflies represent RXTE data simultaneous to 
{\it Beppo}SAX. 
Right low panel: the best sampled BL Lac simultaneous SEDs:
the November 1995 faint state and the great July 1997 flare.}
\label{fig1}
\end{center}
\end{figure*}

{\it Beppo}SAX observed the source while in two different optical states: 
during October--November BL Lac was 1.5-2 times brigher 
than in July (M$_R \sim 14$).
 The differencies are further evident
in the X--ray: when optically faint  the source was not detected 
by the  PDS experiment and
 the [0.6-10] keV LECS--MECS spectrum can be interpreted as inverse Compton
emission, since it is  well fitted 
by a hard power law model of index $\alpha =0.8$ (N$_{\rm H}$ fixed to 
$2.5\times 10^{21}$ cm$^{-2}$; 
Sambruna et al., 1999; Ravasio et al., 2002).\\
During the second observation, 
the PDS detected the source up to $\sim 50$ keV; 
 the [0.3-10] keV  flux was the highest recorded for BL Lac
 and the spectrum was well 
fitted by a convex broken power law, softening at E$_b \sim 2.2$ keV
(see table \ref{tab1}). PDS data lie above the described model
leaving positive residuals becoming larger towards higher energies:
this can be explained as the transition from a steepening synchrotron 
component to an hard inverse Compton, dominating above $\sim 10$ keV.
{\it Beppo}SAX data are confirmed by the simultaneous RXTE [3-15] keV 
spectra: fitting them with power law models, we obtained
$\alpha = 0.88$ and $\alpha=1.45$ 
for the summer and autumn observations respectively (see table \ref{tab1}).\\ 
\begin{table*}[t]
\vspace{0.2cm}
\begin{center}
\begin{tabular}{cccccc}
\hline
Date & Instrument  & $\alpha_1$ & E$_b$ & $\alpha_2$ & F$_{2-10 keV}$ \\
     &           &       & (keV)   &            & (ergs cm$^{-2}$ s$^{-1}$) \\

\hline

\vspace{0.1cm}
26--27/7/2000 & {\it Beppo}SAX & & & $0.81\pm0.07$& $5.81\times10^{-12}$ \\
26/7/2000     & RXTE &  & & $0.88^{+0.93}_{-0.8}$& \\

\hline

\vspace{0.1cm}
31/10--2/11/2000 & {\it Beppo}SAX & $1.455^{+0.08}_{-0.475}$ & $2.23^{+0.73}_{-1.3}$ & $1.65\pm 0.08$ & $2.05\times10^{-11}$ \\

2/11/2000        & RXTE & & &  $1.45^{+0.39}_{-0.35}$ & \\
\hline  
\end{tabular}
\caption{Best fit spectral parameters: N$_{\rm H}$ is fixed at the value
 $2.5\times10^{21}$ cm$^{-2}$. The October--November analysis is performed
only on LECS--MECS data.}
\label{tab1}
\end{center}
\end{table*}
The temporal behaviour of the source confirmed the high state of activity
during the autumnal observation :
{\it Beppo}SAX detected flux variations of more than a factor 3 in 
time scales of 1 hr (see fig. \ref{fig1}).
This is similar to the event flare detected during 
the  observation  of July 1999 (Ravasio et al., 2002). In that occasion, 
the BL Lac X--ray spectrum was displaying 
the transition between the two emission
mechanisms: the flare was visible only in the energy range where synchrotron 
radiation was dominating. In autumn 2000, instead,  
both LECS and MECS were seeing synchrotron emission:
 the fast and large variability is found in the full 0.1--10 keV energy 
 range. 
This extreme behaviour is not surprising since we are observing
the emission of very  energetic electrons ($\gamma \sim 10^5-10^6$)
 that cool very quickly ($\sim 10^3$ sec).
Furthermore, the spectrum is steep: a small change in the spectral slope
will produce large flux variations.\\
\section{Discussion}
The great differencies discussed above are clearly evidenced in the 
Spectral Energy Distributions reported in fig. \ref{fig2}.
To show the complex behaviour of BL Lac
we plotted also two historical simultaneous SEDs, relative to
the faint state of November 1995  
and to the big flare of July 1997:  
the high energy peak of this first SED 
was interpreted by Sambruna et al. (1999) 
as SSC emission (Maraschi, Ghisellini \& Celotti, 1992), while the 
hardness of the  X--to--$\gamma$--ray flaring spectrum was attributed 
by Madejski et al. (1999) to an 
External Compton  mechanism (Sikora, Begelmann \& Rees, 1994).\\
Shifting the synchrotron component in frequencies and fluxes
we can phenomenologically reproduce 
all the observed SEDs, except the one of the big flare of July 1997:
 in that case,
the X--ray Compton spectra was extremely higher (more than a factor 4) 
than all the other  Compton spectra, while the synchrotron 
component was similar. This uniqueness
can be accounted for using a simple scenario: a synchrotron emitting blob
moving along a jet can be inside or outside the Broad Line Region
(see also Ravasio et al., 2002).
If outside, the synchrotron photons will be the only available
targets for inverse Compton scattering; if inside, otherwise, there will be 
also the disk emission reprocessed by the BLR. In the special case
in which a BLR cloud is present along the jet, there would
be a futher target radiation field, composed by the synchrotron radiation
reprocessed by the cloud (Ghisellini \& Madau, 1996). 
This latter case could explain the extraordinary 
X--to--$\gamma$--ray spectra seen during July 1997.
We are not able to distinguish the engine producing the spectra
of 2000, since we lack $\gamma$--ray informations. Anyway the fast variability
suggest the compactness of the emitting region: during the second observation, 
the blob could be inside the BLR, producing an hard high energy component.
\section*{Acknowledgments}
This research was financially supported by the MURST 
and by the Italian Space Agency.


\section*{References}
\begin{thebibliography}{99}

\bibitem{bot} B\"{o}ttcher M. et al., in prep.

\bibitem{bre} Bregman J.N., Glassgold A.E., Huggins P.J. et al., 1990, ApJ, 352, 574

\bibitem{ghi} Ghisellini G. \& Madau P., 1996, MNRAS, 280, 67

\bibitem{kaw} Kawai N., Matsuoka M. et al., 1991, ApJ, 382, 508

\bibitem{mad} Madejski G.M., Sikora M., Jaffe T., Bla\.{z}ejowski M., Jahoda K., Moderski R., 1999, ApJ, 521, 145

\bibitem{man} Mang O. et al., 2001, in proc. of the 27$^{th}$ ICRC, 2658

\bibitem{mar} Maraschi L., Ghisellini G. \& Celotti A., 1992, ApJ, 397, L5

\bibitem{mar} Marscher A.P. et al., in prep.

\bibitem{rav} Ravasio M., Tagliaferri G., Ghisellini G. et al., 2002, A\&A, 383, 763
 
\bibitem{rav2} Ravasio M., Tagliaferri G., Ghisellini G. et al., in prep.

\bibitem{sam} Sambruna R.M., Ghisellini G., Hooper E., Kollgard R.I., Pesce J.E., Urry C.M., 1999, ApJ, 515, 140

\bibitem{sik} Sikora M., Begelman M.C. \& Rees M., 1994, ApJ, 421, 153

\bibitem{vil} Villata M. et al., 2000, A\&A, 363, 108 

\bibitem{vil2} Villata M., Raiteri C.M. et al., 2002, A\&A, 390, 407
\end{thebibliography}
\end{document}